\documentclass[preprint,preprintnumbers,amsmath,amssymb,nofootinbib,aps,usenatbib]{revtex4}
\pdfoutput=1
\usepackage{epsfig}
\usepackage{graphicx}
\usepackage{dcolumn}
\usepackage{bm}
\usepackage{amsmath}
\usepackage{amssymb}
\usepackage{latexsym}
\usepackage{color}
\usepackage{hyperref}
\usepackage{mathrsfs}

\newcommand{\be}{\begin{equation}}
\newcommand{\ee}{\end{equation}}
\newcommand{\bea}{\begin{eqnarray}}
\newcommand{\eea}{\end{eqnarray}}

\newcommand{\like}{\mathscr{L}}
\begin{document}


\title{Dark matter from torsion in Friedmann cosmology}

\author{S. H. Pereira$^{1}$} \email{s.pereira@unesp.br}
\author{A. M. Vicente$^{1}$} \email{amvfisico@gmail.com}
\author{J. F. Jesus$^{1,2}$}\email{jf.jesus@unesp.br}
\author{R. F. L. Holanda$^{3}$} \email{holandarfl@fisica.ufrn.br}

\affiliation{$^1$Universidade Estadual Paulista (UNESP), Faculdade de Engenharia de Guaratinguet\'a, Departamento de F\'isica - Av. Dr. Ariberto Pereira da Cunha 333, 12516-410, Guaratinguet\'a, SP, Brazil
\\
$^2$Universidade Estadual Paulista (UNESP), Instituto de Ci\^encias e Engenharia - R. Geraldo Alckmin, 519, 18409-010, Itapeva, SP, Brazil,
\\
$^3$Departamento de F\'isica Te\'orica e Experimental, Universidade Federal do Rio Grande do Norte, Natal, RN, 59300-000, Brazil
}


\def\zt{\mbox{$z_t$}}

\vspace{1.5cm}
\begin{abstract}
A cosmological model in an Einstein-Cartan framework endowed with torsion is studied. For a torsion function assumed to be proportional to Hubble expansion function, namely $\phi=-\alpha H$, the contribution of torsion function as a dark matter component is studied in two different approaches. In the first one, the total matter energy density is altered by torsion coupling $\alpha$, giving rise to an effective dark matter and cosmological constant terms that reproduce quite well the flat cosmic concordance model. In the second approach, starting with just standard baryonic matter plus a cosmological constant term, it is obtained that the coupling of torsion with baryons and cosmological constant term naturally gives rise to a dark matter contribution, together a modified cosmological term. In this model the dark matter sector can be interpreted as an effective coupling of the torsion function with the ordinary baryonic matter and cosmological constant. Finally, it is shown that both models are totally compatible with recent cosmological data from Supernovae and Hubble parameter measurements. 

\end{abstract}

\maketitle



\section{Introduction}
\label{sec:introd}

It is well known that the standard model of cosmology, known as flat $\Lambda$CDM model,  predicts quite well the overall evolution of the Universe, from long time before the radiation era up the current acceleration phase. The model has been tested over several cosmological and astrophysical observations, having the general relativity (GR) as the theoretical background for the model. However, there exist some specific observational discrepancies at both small scales and large scales that open the possibility for extensions of the standard model. We cite the small-scale problems \cite{Peebles21}, the cosmic curvature problem \cite{DiValentino:2019qzk}, the Hubble tension \cite{Martinelli:2019krf,valentino2021}, among others\footnote{See Ref.~\cite{Bull:2015stt} for a recent review on the standard $\Lambda$CDM problems}. 

A direct extension of GR that alters the dynamics of expansion of the universe is based on the Einstein-Cartan-Kibble-Sciama (ECKS) gravitational theory \cite{Sciama:1964wt,Hehl:1971qi,Hehl:1976kj,Shapiro:2001rz}, where the spin tensor of matter act as the source of torsion, generalizing the GR equations \cite{Pasmat2017}. The generalized Friedamm equations that follows from this approach have been studied recently in different contexts. Dark energy effects and late time cosmology have been addressed in \cite{Ivanov,Kranas2019,Barrow2019,Pereira2019,Cruz2020}. Inflationary models were studied in \cite{Popla,Thiago2020}, and torsion as an alternative to dark matter were explored in \cite{Tilquin,Marques2020}. A recent review on Einstein-Cartan cosmologies was done by Medina et. al. \cite{Medina2019}.

In the present paper we study the torsion effects in a Friedmann cosmology as a candidate to dark matter in the universe. After review and constraint the free parameters of a model where torsion just alters the presence of a given matter energy density plus a cosmological term \cite{Marques2020}, we propose a new scenery where the only matter content is baryonic matter plus a cosmological term. We show that the coupling of the torsion with baryonic and cosmological constant terms naturally leads to an effective dark matter contribution. The cosmological constant term is also modified by the torsion coupling, resulting in an effective cosmological constant. The free parameters of the model are constrained by observational data of Supernovae and Hubble parameter measurements. 

In Section II we present the main equations of the model. Section III we present the constraints with observational data and analysis. Section IV contain the conclusions. We left to a brief Appendix the ECKS equations, from which follows the starting point equations of Section II.

\section{Friedmann cosmology with torsion}
\label{sec:torsion}

We follow the same development and notation from \cite{Kranas2019}, where the torsion field is represented by $\phi(t)$. The Friedmann equations including a general matter density $\rho$, pressure $p$ cosmological constant $\Lambda$ and curvature $k$ are (see Appendix A for a brief overview):
    \begin{equation}
        H^2 =\frac{8\pi G}{3}\rho  + \frac{\Lambda}{3}- \frac{k}{a^2}-4\phi^2 - 4H\phi\,,\label{H2}
    \end{equation}
    \begin{equation}
         \dot{H}+H^2 =-\frac{4\pi G}{3}(\rho+3p)  + \frac{\Lambda}{3} - 2\dot{\phi} - 2H\phi\,,\label{Hd2}
    \end{equation}
   where $H=\dot{a}/a$ is the Hubble function and $a(t)$ the scale factor of the universe. For a barotropic matter satisfying an equation of state of the form $p=w \rho$, the continuity equation reads:
    \begin{equation}
        \dot{\rho}+3(1+w)H\rho + 2(1+3w)\phi \rho = 4\phi\frac{\Lambda}{8\pi G}\,.\label{rhodot}
    \end{equation}
Given a torsion function $\phi(t)$ the above system of equations can be solved, at least numerically. Next we will analyse two different approaches where the torsion function is just proportional to the Hubble parameter $H$.

For the specific choice $\phi(t) = -\alpha H(t)$, (or  $\phi(t) = \lambda H(t)$ as considered by \cite{Kranas2019,Marques2020}), where $\alpha$ (or $\lambda$) is a constant that characterises the strength of torsion field, the model is known as steady-state torsion.

Now we will analyse this specific model by two different approaches. The Friedmann equation (\ref{H2}) turns:
 \begin{equation}
        H^2 =\frac{1}{(1-4\alpha+4\alpha^2)}\bigg[\frac{8\pi G}{3}\rho  + \frac{\Lambda}{3} - \frac{k}{a^2}\bigg]\,,\label{H2a}
\end{equation}
and the solution for the energy density from (\ref{rhodot}) is:
\begin{eqnarray}
    \rho(a) &=& \Bigg[\rho_0 + \frac{\alpha \Lambda}{2\pi G [3(1+w)-2\alpha(1+3w)]}\Bigg]\Bigg(\frac{a_0}{a} \Bigg)^{3(1+w)-2\alpha(1+3w)}\nonumber\\
    &&- \frac{\alpha \Lambda}{2\pi G [3(1+w)-2\alpha(1+3w)]}\,.\label{sol2}
\end{eqnarray}
This solution warranty that for the present time $a=a_0$ we have $\rho(a_0) = \rho_0$, which shows that energy density is affected by the presence of both torsion and cosmological constant term along evolution, but for the present time it is $\rho_0$. For $\alpha = 0$ and $\Lambda \neq0$ the evolution of the energy density is exactly the expected one, namely $ \sim a^{-3(1+w)}$. However, for $\alpha \neq 0$ and $\Lambda =0$ the evolution of matter energy density if affected by torsion solely through the exponent in $a$. The second term also will be dominant in the future, when $a >> a_0$, if $2\alpha(1+3w)<3(1+w)$.

By assuming that the matter content is of dust type (dark matter or baryonic matter), we take $w=0$ and the energy density is:
\begin{eqnarray}
    \rho_m(a) = \Bigg[\rho_{m0} + \frac{\alpha \Lambda}{2\pi G (3-2\alpha)}\Bigg]\Bigg(\frac{a_0}{a} \Bigg)^{3-2\alpha} - \frac{\alpha \Lambda}{2\pi G (3-2\alpha)}\,.\label{sol22}
\end{eqnarray}

In which follows we will analyze the model described by (\ref{H2a}) with the solution (\ref{sol22}) in two different approaches. The analysis will be done in a flat background $(k=0)$ in order to compare the results to the ones of standard model of cosmology, namely the flat $\Lambda$CDM model.  

\subsection{Case I}

In the first case we take (\ref{H2a}) with $k=0$  and substitute the energy density (\ref{sol22}). We aims to constraint the free parameters of the model, namely $\alpha$ and the matter density parameter. This model has been analyzed in \cite{Marques2020}, in different contexts, including a general equation of state parameter $w$.

By defining the present density parameters\footnote{We are omitting the subscript '0' to represent present day values of density parameters just for short.} $\Omega_{m} = \frac{8\pi G \rho_{m}}{3H_0^2}$ and $\Omega_{\Lambda}=\frac{\Lambda}{3H_0^2}$, the Friedmann equation (\ref{H2a}) can be written as:
\begin{equation}
    \frac{H^2}{H_0^2} = \frac{1}{(1-2\alpha)^2}\Bigg[\Big(\Omega_{m}+\frac{4\alpha}{(3-2\alpha)}\Omega_\Lambda\Big)(1+z)^{3-2\alpha} + \Omega_{\Lambda}\Big(1-\frac{4\alpha}{3-2\alpha}\Big)\Bigg]\,.\label{H2c1a}
\end{equation}
By using the Friedmann constraint at $z=0$, namely $1=(\Omega_m + \Omega_{\Lambda})/(1-2\alpha)^2$, we can express $\Omega_\Lambda $ as a function of $\Omega_{m}$, namely $\Omega_\Lambda = (1-2\alpha)^2 - \Omega_m$. Thus, leaving $H_0$ to be a free parameter, we have a model with 3 free parameters, namely $H_0$, $\alpha$ and $\Omega_{m}$. Notice that if we define the new parameters:
\begin{equation}
\Omega_{dm} = \frac{1}{(1-2\alpha)^2}\left[\Omega_{m}+\frac{4\alpha}{(3-2\alpha)}\Omega_\Lambda\right] \label{Eq8}
\end{equation}
and 
\begin{equation}
\Omega_{\Lambda\, eff} = \frac{\Omega_\Lambda}{(1-2\alpha)^2}\left[1-\frac{4\alpha}{(3-2\alpha)}\right] \label{Eq9}
\end{equation}
as effective dark matter and cosmological constant density parameters, we have the constraint for the present day ($z=0$):
\begin{equation}
    1 = \Omega_{dm} + \Omega_{\Lambda\, eff}\,,\label{H2c1b}
\end{equation}
which can be directly compared to the $\Lambda$CDM model after the constraints of $\Omega_{dm}$. This analysis will be done in next Section.

\subsection{Case II}

 Here we make a quite different analysis. Starting from (\ref{H2a}) with $k=0$ and assuming $\alpha$ being a small parameter (to be verified later with observational constraint), we call $X = 4\alpha - 4\alpha^2$ and make an expansion:
\begin{equation}
    \frac{1}{1-X} = 1 + X + X^2 + X^3 + \dots = 1 + f(\alpha)\,, \label{X}
\end{equation}
which defines $f(\alpha) = X + X^2 + X^3 + \dots$, or
\begin{equation}
    f(\alpha) = \frac{1}{1-X} - 1 = \frac{4\alpha(1-\alpha)}{1-4\alpha+4\alpha^2}\,. \label{f}
\end{equation}
For $|X| < 1$ the geometric series (\ref{X}) converges, which puts a limit on $\alpha$ parameter, namely:
\begin{equation}
    \frac{1}{2}(1-\sqrt{2}) < \alpha < \frac{1}{2}\,,\hspace{0.8cm}\frac{1}{2} < \alpha < \frac{1}{2}(1+\sqrt{2})\label{alpha}
\end{equation}
which is equivalent to about $-0.207 \lesssim \alpha < 0.5$ and $0.5 < \alpha \lesssim +1.207$. For $\alpha = \frac{1}{2}$ the denominator of (\ref{f}) diverges. 

Using (\ref{X}) into (\ref{H2a}) we obtain:
 \begin{equation}
        H^2 =\bigg[\frac{8\pi G}{3}\rho  + \frac{\Lambda}{3}\bigg] + f(\alpha) \bigg[\frac{8\pi G}{3}\rho  + \frac{\Lambda}{3} \bigg]\,.\label{H2aa}
\end{equation}
The first term on r.h.s. is equivalent to flat $\Lambda$CDM model for a given a matter density $\rho$, while the second term represents all the $\Lambda$CDM components coupled to torsion through $f(\alpha)$. For $\alpha \to 0$ we recover exactly the $\Lambda$CDM model. 

Now we make the particular choice that only matter content is the standard baryonic matter, $\rho_m = \rho_b$ (contrary to last case, where $\rho$ were taken as the total matter density). By using (\ref{sol22}) into (\ref{H2aa}) we have:
\begin{eqnarray}
    \frac{H^2}{H_0^2} &=& \Omega_{b0}(1+f)(1+z)^{3-2\alpha} + \Omega_{\Lambda}(1+f)\frac{4\alpha}{3-2\alpha}(1+z)^{3-2\alpha} \nonumber\\
    && +  \Omega_{\Lambda}(1+f)\bigg(1-\frac{4\alpha}{3-2\alpha}\bigg)\,,\label{mod42}
\end{eqnarray}
with $f(\alpha)$ given by (\ref{f}). For the present time ($z=0$) we have that $\Omega_\Lambda$ can be written as a function of $\Omega_{b0}$ and $\alpha$ (through $f$) as  $\Omega_\Lambda= 1/(1+f) - \Omega_{b0}$. By fixing $\Omega_{b0}$ we are left with a two parameter model, namely $H_0$ and $\alpha$.

Now we explicitly separate out the pure baryonic density term from the terms depending on torsion:
\begin{eqnarray}
    \frac{H^2}{H_0^2} &=& \Omega_{b0}(1+z)^{3-2\alpha} + \bigg[f\Omega_{b0} + \Omega_{\Lambda}(1+f)\frac{4\alpha}{3-2\alpha}\bigg](1+z)^{3-2\alpha} \nonumber\\
    && +  \Omega_{\Lambda}(1+f)\bigg(1-\frac{4\alpha}{3-2\alpha}\bigg)\,.\label{mod42c}
\end{eqnarray}
By defining the new density parameters:
\begin{equation}
    \Omega_{dm}=f\Omega_{b0} + \Omega_{\Lambda}(1+f)\frac{4\alpha}{3-2\alpha}\,\label{42c1}
\end{equation}
\begin{equation}
    \Omega_{\Lambda eff}=\Omega_{\Lambda}(1+f)\bigg(1-\frac{4\alpha}{3-2\alpha}\bigg)\,,\label{42c2}
\end{equation}
as effective dark matter and cosmological constant density parameters, the model (\ref{mod42c}) can be written as:
\begin{eqnarray}
    \frac{H^2}{H_0^2} &=& \Omega_{b0}(1+z)^{3-2\alpha} + \Omega_{dm}(1+z)^{3-2\alpha} + \Omega_{\Lambda eff}\,.\label{mod42d}
\end{eqnarray}
For the present time ($z=0$) we have the constraint:
 \begin{equation}
    1 = \Omega_{b0}+ \Omega_{dm}+\Omega_{\Lambda eff}\,,\label{mod42a}
\end{equation}
which can be compared to the flat $\Lambda$CDM model with the explicit term of baryonic matter separate from dark matter one.

Equation (\ref{mod42d}) along with (\ref{42c1}) and (\ref{42c2}) contains the main result of the present work. Notice that written in the form (\ref{mod42d}), the effective dark matter component is just a combination of the coupling of the torsion function (characterized by $\alpha$) with the baryonic density parameter and the standard cosmological constant term. In this sense, the effect of dark matter can be seen as the net effect of torsion coupling with baryonic matter and cosmological constant. Finally, the standard cosmological constant term $\Omega_\Lambda$ is also affected by the torsion coupling, producing an effective cosmological constant term. When $\alpha \to 0$ the standard model is recovered, with just the presence of baryonic matter and cosmological constant.

Next we will analyze the Cases I and II separately, the first one  just for comparison with the results of \cite{Marques2020}, and the second case contains the main results of the paper.

\section{Constraints with observational data and analysis}

Now we will analyze the model and constraint the free parameters by using Hubble parameter data,  $H(z)$, and Supernovae Type Ia (SNe Ia) data. The 51 $H(z)$ data compilation used is grouped in Ref. \cite{Magana2018}, consisting of 20 clustering (obtained from measurements of peaks of baryonic acoustic oscillations and through
correlation function of luminous red galaxies) and 31 differential age $H(z)$ data, known as Cosmic Chronometers. The first set of $H(z)$ data are model dependent, based on $\Lambda$CDM model, while the second one is model independent. The $H(z)$ data cover the redshift  range $0.07 < z < 2.36$. For Supernovae we consider the Pantheon sample \cite{pantheon}, one of the largest combined sample of SNe Ia, consisting of a total of 1048 SNe Ia in the range $0.01 < z < 2.3$. Pantheon sample uses a method of calibration with bias corrections, which allows to determine SNe Ia
distances without the necessity to fit the Supernovae parameters jointly with cosmological parameters. Thus, Pantheon provide corrected estimates of overall normalization flux $m_B$ in order to constrain the cosmological parameters. 

We have used flat priors for all parameters. We determine the best-fit values and uncertainty of the parameters by maximizing the likelihood
function. For $H(z)$, the likelihood distribution function is $\like_H \propto e^{-\frac{\chi^2_H}{2}}$, where:
\begin{equation}
\chi^2_H = \sum_{i = 1}^{51}\frac{{\left[ H_{\mathrm{obs},i} - H(z_i,\mathbf{p})\right] }^{2}}{\sigma^{2}_{H_\mathrm{obs},i}},
\label{chi2H}
\end{equation}
 $\mathbf{p}$ is the vector of free parameters of the model. 

For the SNe Ia data represented by the Pantheon sample, the likelihood function is $\like_{SN} \propto e^{-\frac{\chi^2_{SN}}{2}}$, where:
\be
\chi^2_{SN}=\mathbf{\Delta m}^T\cdot\mathbf{C}^{-1}\cdot\mathbf{\Delta m},
\ee
where $\textbf{C}$ is a covariance matrix for the
parameters including statistical and systematic uncertainties \cite{Betoule}, $\mathbf{\Delta m}=m_B-5\log_{10}D_L(z,\mathbf{p})+\mathcal{M}$, where $D_L$ is the luminous distance for the flat background given by:
\begin{equation}
\label{eq:DM}
  D_L(z,\mathbf{p}) = 
   \frac{(1+z)}{H_0} \int^z_0 \frac{H_0}{H(z',\mathbf{p})}dz'.
\end{equation}
where $\mathcal{M}$ is a nuisance parameter which contains $H_0$. We choose to project over $\mathcal{M}$, thus we find the projected $\chi^2_{\mathrm{SNproj}}$:
\be
\chi^2_\mathrm{SNproj}=S_{mm}-\frac{S_m^2}{S_A},
\ee
where $S_{mm}=\sum_{i,j}\Delta m_i\Delta m_j(C^{-1})_{ij}=\mathbf{\Delta m}^T\cdot\mathbf{C}^{-1}\cdot\mathbf{\Delta m}$, $S_m=\sum_{i,j}\Delta m_i(C^{-1})_{ij}=\mathbf{\Delta m}^T\cdot\mathbf{C}^{-1}\cdot\mathbf{1}$ and $S_A=\sum_{i,j}(C^{-1})_{ij}=\mathbf{1}^T\cdot\mathbf{C}^{-1}\cdot\mathbf{1}$.

The constraints over the free parameters are obtained by sampling the combined likelihood function $\like \propto e^{-\frac{1}{2}(\chi^2_{H}+\chi^2_{SNproj})}$ through the Affine Invariant method of Monte Carlo Markov Chain (MCMC) analysis implemented in {\sffamily Python} language by using {\sffamily emcee} software. (See  \cite{GoodmanWeare,ForemanMackey13} for further details).


\subsection{Case I}

The primary model, Case I, is described by Eq.  (\ref{H2c1a}), but remember that it comes directly from (\ref{H2a}). For this case, if $\rho = \rho_m$ (the total matter content) and we assume $\alpha << 1$ (to be verified by observational constraint), we see that the model is almost equivalent to the standard $\Lambda$CDM model. For this reason here we have used the complete 51 $H(z)$ data plus the SN Ia - Phanteon sample. This choice also permits a more direct comparison with the results of \cite{Marques2020}, where it was used 38 $H(z)$ data \cite{Farooq38} which also includes model dependent measurements.

\begin{figure}[t!]
\centering
\includegraphics[width=0.8\linewidth]{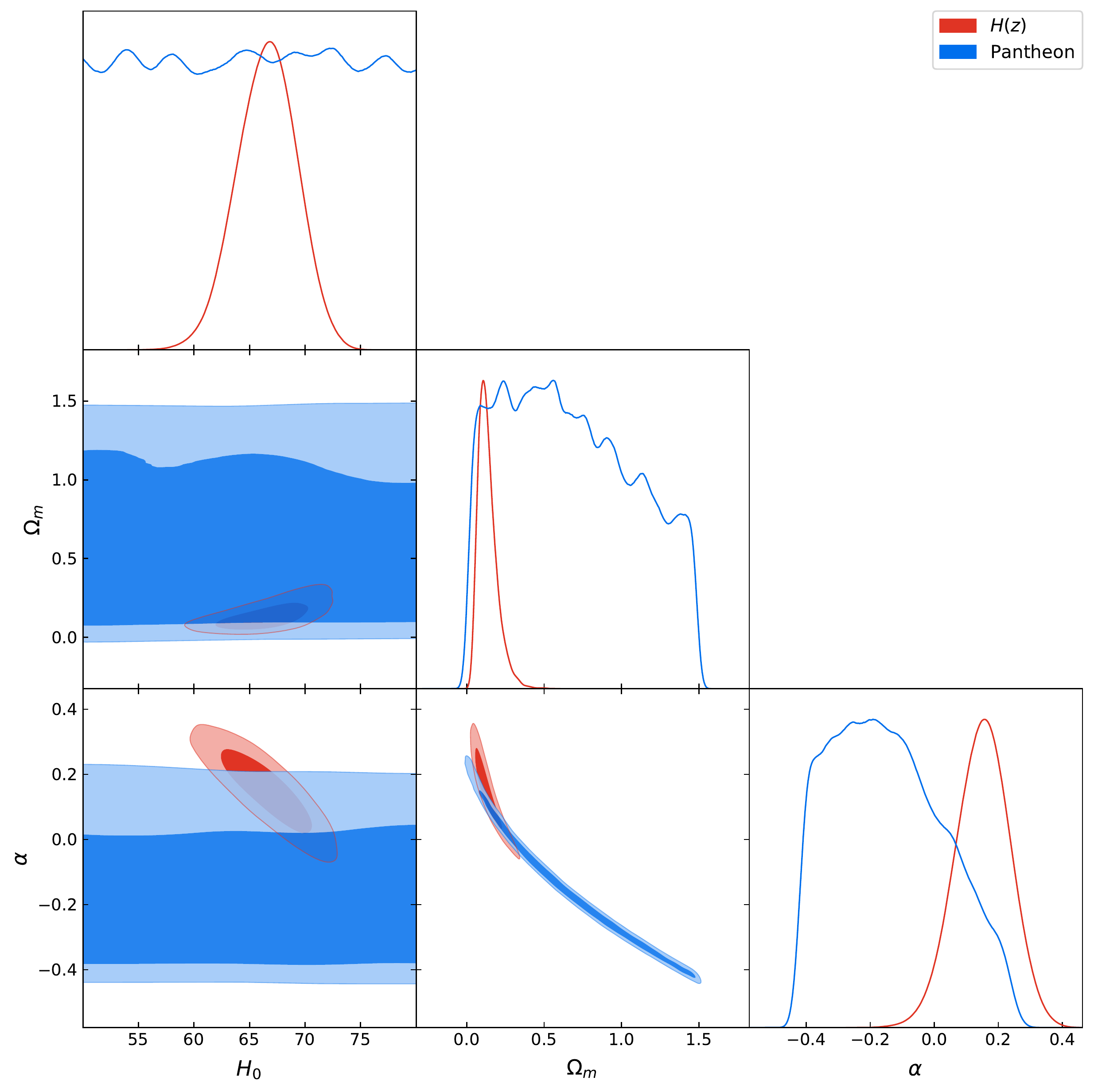}
\caption{Separate analysis using SNe Ia - Pantheon (blue) and 51 $H(z)$ data  (red), for the constraints of the main parameters,  $H_0$, $\Omega_m$ and $\alpha$, for Case I.}
\label{fig01}
\end{figure}
\begin{figure}[h!]
\centering
\includegraphics[width=0.8\linewidth]{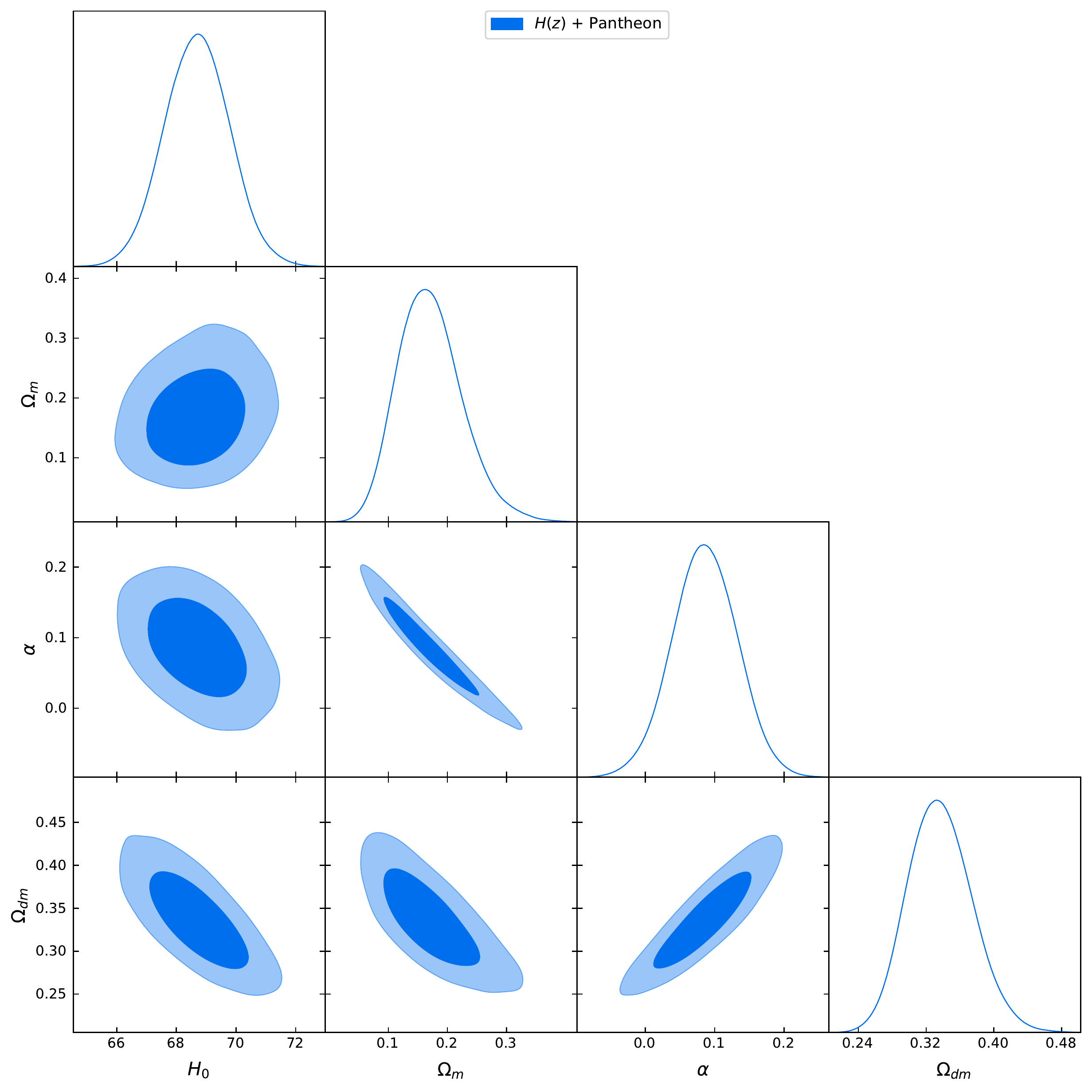}
\caption{Joint analysis using SNe Ia - Pantheon and 51 $H(z)$ data, for the constraints of the main parameters, $H_0$, $\Omega_m$ and $\alpha$ and for the derived parameter  $\Omega_{dm}$, for Case I.}
\label{fig02}
\end{figure}

The free parameters of the model are $\mathbf{p}=[H_0,\,\Omega_m,\alpha]$.  The $H(z)$ (red) and SNe Ia-Pantheon (blue) one-dimensional likelihoods and two-dimensional confidence contours for the free cosmological parameters are shown in Figure \ref{fig01}, which shows that two different set of parameters can be combined. This is shown in Figure \ref{fig02}, where we have also added the contour for the derived parameter $\Omega_{dm}$ from (\ref{Eq8}). The mean values of the parameters and 95\% c.l. are given in Table \ref{tab1}. Notice that the mean value of $H_0$ is in good agreement to latest Planck 2018 results \cite{Planck2018} ($H_0=67.36\pm 0.54$). The value for the $\alpha$ parameter is also in agreement to the one obtained in \cite{Marques2020} ($\lambda = -0.07^{+0.05}_{-0.04}$, notice that $\lambda = -\alpha$ when compare \cite{Marques2020} to the present work). Notice that the supposition $\alpha << 1$ is satisfied even at $2\sigma$c.l.. The value of $\Omega_m$ obtained from \cite{Marques2020} ($\Omega_m = 0.18^{+0.06}_{-0.03}$) is also compatible to our result. A much more interesting result that we have obtained is by using the definition of an effective dark matter component, $\Omega_{dm}$, given by (\ref{Eq8}). With such definition the value obtained in our analysis, $\Omega_{dm}=0.336^{+0.078}_{-0.070}$ is in good agreement to Planck 2018 results \cite{Planck2018} ($\Omega_{dm}=0.315\pm{0.007}$).

\begin{table}[t!]
    \centering
\begin{tabular} { l  c}

 Parameter &  95\% limits\\
\hline
{\boldmath$H_0            $} & $68.7^{+2.2}_{-2.2}        $\\

{\boldmath$\Omega_{m}     $} & $0.17^{+0.11}_{-0.10}      $\\

{\boldmath$\alpha         $} & $0.086^{+0.094}_{-0.095}   $\\

$\Omega_{dm}               $ & $0.337^{+0.080}_{-0.070}   $\\
\hline
\end{tabular}
\caption{Mean values and 95\% c.l. for the parameters $H_0$, $\Omega_m$, $\alpha$ and $\Omega_{dm}$ for Case I.}
    \label{tab1}
\end{table}

These results for the Case I show that torsion contribution can be responsible to explain correctly the present day values of the main cosmological parameters, reproducing a total dark matter content in full agreement to standard model results. Notice, however, that in this primary model we can not separate the baryonic contribution from the dark matter one. All we obtain is an effective dark matter contribution. The explicit inclusion of baryonic matter will be done in the next Case, which represents the main results of the present work.

\subsection{Case II}

For the Case II, we start from (\ref{mod42}), written $\Omega_\Lambda$ as a function of $\Omega_{b0}$. By fixing $\Omega_{b0}$ we have a two parameter model, namely $H_0$ and $\alpha$ (through $f$ given by (\ref{f})). Thus $\mathbf{p}=[H_0,\alpha]$. However, we do not have a fixed value for the baryonic density parameter $\Omega_{b0}$ itself. Thus, we use the BBN constraint on the baryion density, namely $\Omega_b h^2 = 0.022353 \pm 0.00033$, with $h \equiv H/100$, and the present day value of the baryon density parameter can be obtained from $\Omega_{b0} (H_0/100)^2 \simeq 0.022353$, since that $H_0$ is also a free parameter in our model. Since the matter content assumed is just the baryonic matter, this model is quite different from $\Lambda$CDM model at this stage, and we can not assume $\alpha << 1$. Thus we use just the 31 model independent $H(z)$ data - Cosmic Chronometers, plus SNe Ia - Pantheon sample.

\begin{figure}[h!]
\centering
\includegraphics[width=0.7\linewidth]{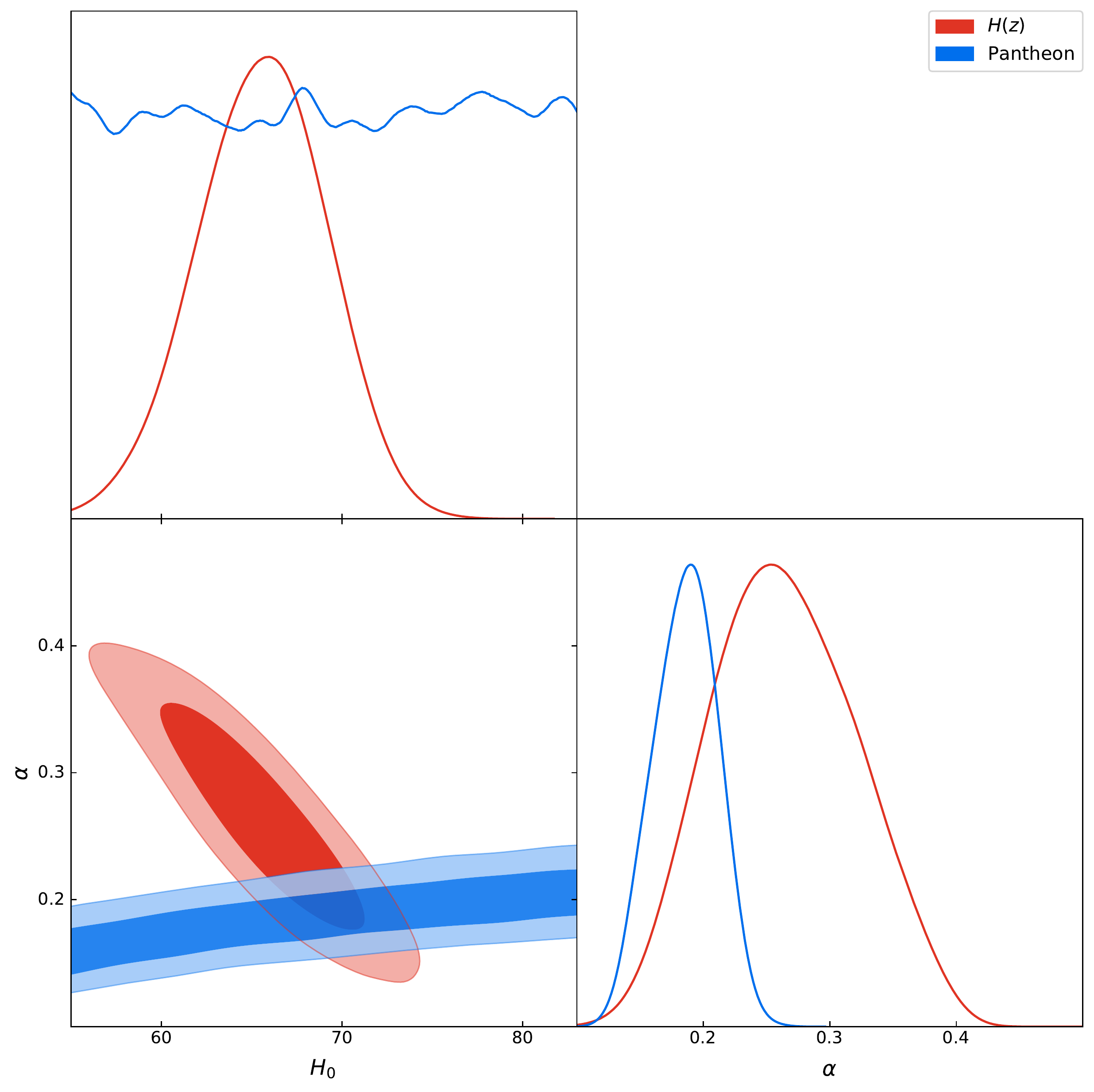}
\caption{Separate analysis using SNe Ia - Pantheon (blue) and 31 $H(z)$ data - Cosmic Chronometers (red), for the constraints of the main parameters,  $H_0$ and $\alpha$, for Case II.}
\label{fig03}
\end{figure}

\begin{figure}[h!]
\centering
\includegraphics[width=0.8\linewidth]{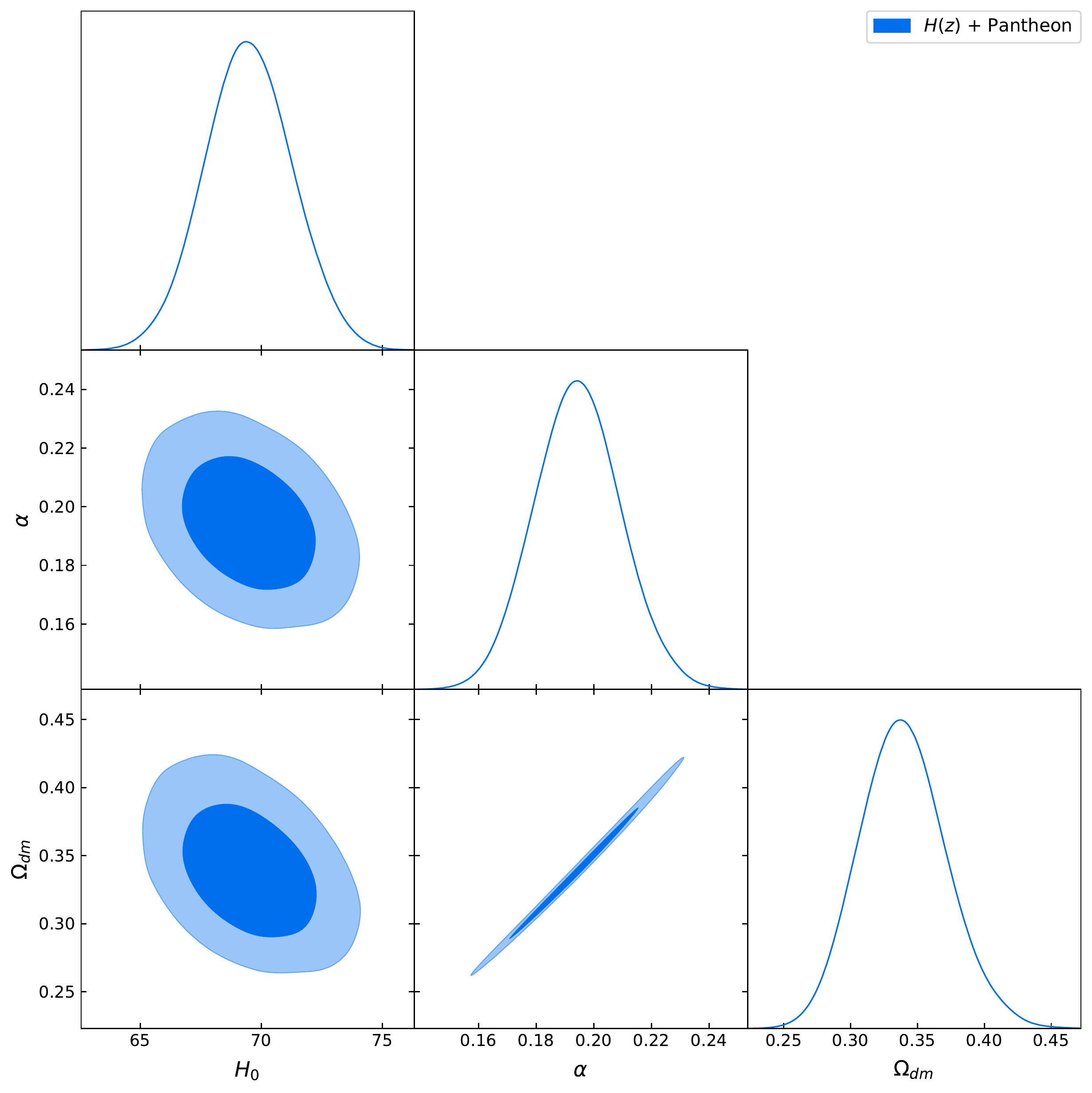}
\caption{Joint analysis using SNe Ia - Pantheon and 31 $H(z)$ data - Cosmic Chronometers, for the constraints of the main parameters,  $H_0$ and $\alpha$, and also for the derived parameter $\Omega_{dm}$, for Case II.}
\label{fig04}
\end{figure}

Figure \ref{fig03} shows the contours for the parameters $H_0$ and $\alpha$ for separate analysis using SNe Ia - Pantheon (blue) and 31 $H(z)$ - Cosmic Chronometers (red) data, at $1\sigma$ and $2\sigma$ c.l.. We see that the joint analysis is possible, which is shown in Figure \ref{fig04}. The contours for the derived parameter of effective dark matter parameter obtained from (\ref{42c1}) is also present. Table \ref{tab2} shows the mean values for the main parameters $H_0$ and $\alpha$ at $95\%$ c.l., and also the value obtained for the effective dark matter parameter $\Omega_{dm}$. 

First, notice that the mean value for $\alpha$ satisfies the condition (\ref{alpha}), a necessary condition to the convergence of the series (\ref{X}), the cornerstone of the model. Only under this condition the $f$ function can be put as a multiplicative factor in (\ref{H2aa}). Second, the value of $H_0$ obtained is in full agreement to latest Planck 2018 results. The value for the effective dark matter density parameter, $\Omega_{dm}$, is also in good agreement to Planck 2018 results, even though the model be quite different from $\Lambda$CDM model, the constraint (\ref{mod42a}) for present time must be satisfied with the effective parameters.

The analysis of this second case shows that a model starting with just baryonic matter plus a cosmological constant term, in the presence of a torsion function described by $\phi = -\alpha H$, correctly reproduces an universe with an additional term of dark matter. This last term comes naturally from the coupling of the torsion parameter $\alpha$ to baryonic and cosmological constant terms.

\begin{table}[h!]
    \centering
\begin{tabular} { l  c}

 Parameter &  95\% limits\\
\hline
{\boldmath$H_0            $} & $69.5^{+3.7}_{-3.6}        $\\

{\boldmath$\alpha         $} & $0.194^{+0.031}_{-0.029}   $\\

$\Omega_{dm}               $ & $0.339^{+0.069}_{-0.062}   $\\

\hline
\end{tabular}

    \caption{Mean values and 95\% c.l. for the parameters, $H_0$, $\alpha$ and $\Omega_{dm}$.}
    \label{tab2}
\end{table}

\section{Concluding remarks}

We have studied the torsion effects in cosmology as a candidate to dark matter in the universe. The torsion function considered here was of the type $\phi = -\alpha H$. The free parameters of the model were constrained by observational data of Supernovae and Hubble parameter measurements.

In the first case, already studied in \cite{Marques2020}, we just made a new interpretation for the effective dark matter that appears coupled to torsion parameter $\alpha$, obtaining $\Omega_{dm} = 0.337^{+0.080}_{-0.070}$, $H_0 = 68.7\pm 2.2$ and $\alpha = 0.086^{+0.094}_{-0.095}$ at 95\% c.l.. In the second case we started with just standard baryonic matter plus a cosmological constant term and show that the coupling of the torsion with baryonic and cosmological terms naturally leads to an effective dark matter contribution, giving $\Omega_{dm} = 0.339^{+0.069}_{-0.062}$, $H_0 = 69.5^{+3.7}_{-3.6}$ and $\alpha = 0.194^{+0.031}_{-0.029}$ at 95\% c.l.. Although being a model lightly different from $\Lambda$CDM model, the values of the dark matter density parameter and $H_0$ obtained in both cases are in full agreement to latest Planck 2018 results \cite{Planck2018}. The physical mechanism for the appearing of the dark matter in the second case is much more interesting, since that it appears due to a natural coupling of the torsion parameter to baryonic and cosmological constant terms. In this sense, dark matter can be interpreted as the effect of torsion around standard matter and cosmological constant.

As a final comment, let us recall that in \cite{Kranas2019} the torsion effect on the
primordial nucleosynthesis of helium-4 was studied, and a narrow interval for the $\lambda = -\alpha$ parameter was found $(-0.0058 < \lambda < +0.0194)$. However the analysis was done with vanishing cosmological constant ($\Lambda = 0$) in a flat and radiation dominated universe, thus a direct comparison with the value obtained here must be avoided. Also, the same mechanism of coupling of torsion function with standard baryonic matter will also act on the radiation field, thus we expect the appearing of a kind of dark radiation in the model. Such contribution would also be relevant in computing the freeze-out temperature of the particles at kinetic equilibrium in primordial nucleosynthesis. 

\appendix
\section{The ECKS equations}

The Einstein-Cartan-Kibble-Sciama equations are briefly presented here. We follow the same notation of \cite{Kranas2019}. The equations of gravitation in ECKS framework maintain the same form as the standard one in terms of Ricci tensor, Ricci scalar and energy momentum tensor, namely:
\begin{equation}
    R_{\mu\nu} - \frac{1}{2}R g_{\mu\nu}=\kappa T_{\mu\nu} - \Lambda g_{\mu\nu}\,,\label{Rmunu}
\end{equation}
with $\kappa = {8\pi G}$. However the affine connection is endowed with an antisymmetric part due to torsion, namely $\Gamma^\alpha_{~~\mu\nu}=\Tilde{\Gamma}^\alpha_{~~\mu\nu}+K^\alpha_{~~\mu\nu}$, where $\Tilde{\Gamma}^\alpha_{~~\mu\nu}$ defines the symmetric Christoffel symbols and $K^\alpha_{~~\mu\nu}$ defines the contorsion tensor written in terms of the torsion tensor $S^\alpha_{~~\mu\nu}$,
\begin{equation}
    K_{\alpha\mu\nu} = S_{\alpha\mu\nu} + S_{\mu\nu\alpha} + S_{\nu\mu\alpha} = S_{\alpha\mu\nu} + 2S_{(\mu\nu)\alpha}\,.\label{K}
\end{equation}
The torsion tensor satisfies $S^\alpha_{~~\mu\nu} = - S^\alpha_{~~\nu\mu}$, thus $\Gamma^\alpha_{~~\mu\nu}=\Gamma^\alpha_{~~(\mu\nu)}+S^\alpha_{~~\mu\nu}$ and $\Gamma^\alpha_{~~(\mu\nu)} = \Tilde{\Gamma}^\alpha_{~~\mu\nu} + 2S_{(\mu\nu)}^{~~\alpha}$. 

In a homogeneous and isotropic Friedmann background they are given by \cite{Kranas2019}:
\begin{equation}
  S_\alpha = -3\phi u_\alpha \hspace{1cm} S_{\alpha\mu\nu} = \phi (h_{\alpha\mu} u_\nu - h_{\alpha\nu}u_\mu)  \,, \label{SS}
\end{equation}
where $\phi=\phi(t)$ is a time dependent function representing torsion contribution due to homogeneity of space,  $h_{\mu\nu}$ is a projection tensor, symmetric and orthogonal to the 4-vector velocity $u_\mu$. Thus:
\begin{equation}
    S^i_{~0j} = -S^i_{~j0} = \phi.
\end{equation}

The Ricci tensor is written in the usual form:
\begin{equation}
    R_{\mu\nu} = -\partial_\nu\Gamma^\alpha_{~\mu\alpha} + \partial_\alpha\Gamma^\alpha_{~\mu\nu} - \Gamma^\beta_{~\mu\alpha}\Gamma^\alpha_{~\beta\nu} + \Gamma^\beta_{~\mu\nu}\Gamma^\alpha_{~\beta\alpha}\,,
\end{equation}
and for the ordinary matter satisfying an energy-momentum tensor of a perfect fluid:
\begin{equation}
    T_{\mu\nu} = \rho u_\mu u_\nu + p h_{\mu\nu}\,,
\end{equation}
the Friedmann equations that follows from (\ref{Rmunu}) are given by (\ref{H2})-(\ref{Hd2}). See Appendix B of \cite{Kranas2019} for a detailed derivation.

\begin{acknowledgments}
SHP acknowledges financial support from  {Conselho Nacional de Desenvolvimento Cient\'ifico e Tecnol\'ogico} (CNPq)  (No. 303583/2018-5). This study was financed in part by the Coordena\c{c}\~ao de Aperfei\c{c}oamento de Pessoal de N\'ivel Superior - Brasil (CAPES) - Finance Code 001.
\end{acknowledgments}


\end{document}